\documentclass[prl,twocolumn]{revtex4-1}
\usepackage{times}

\def\_#1{{\bf #1}}

\def\.{\cdot}

\def\=#1{\overline{\overline #1}}

\begin{document}

{\large\bf Comment on ''Electromagnetic Radiation under Explicit Symmetry Breaking''\\ PRL 114, 147701 (2015)}
\medskip

Recently published paper~\cite{PRL} contains several misleading statements and misinterpretations of known facts. The main massage of the paper~\cite{PRL} is as follows: ``\textit{We have shown that explicit symmetry breaking in the structural configuration of charges leads to symmetry breaking of the electric field which results in electromagnetic radiation due to non-conservative current within a localized region of space and time}" seems to transcend mere empiricism, touching the theoretical foundations of electromagnetism. Moreover, basic mistakes are numerous in this article and its main claim is wrong. Below we prove it citing the paper and arguing against it.

\textbf{Abstract:}

1. "\textit{We report our observation that radiation from a system of accelerating charges is possible only when there is explicit breaking of symmetry in the electric field in space within the spatial configuration of the radiating system}" -- This observation (G.G. Thomson) is more than 120 years old, see e.g. \cite{1} or any other tutorial of general physics, chapter "Electricity and magnetism"

2. "…\textit{current within an enclosed area around the radiating structure is not conserved.}" -- The law of current conservation does not exist. Only the steady current conserves and radiating currents does not conserve (neither within an area nor in time), see e.g. \cite{1} or any other tutorial of general physics, volume "Electricity and magnetism".

3. "\textit{Finally, it is argued that symmetry of a resonator of any form can be explicitly broken to create a radiating antenna.}" -- This assertion is wrong. The theory of radiation from resonators is well elaborated, including the case of dielectric resonators, see e.g. in \cite{2}.  The radiation from resonators has nothing to do with symmetry or asymmetry of the resonator and results from their finite resonance quality, implying the radiation losses. Recently, the radiation of optical resonators symmetrically or asymmetrically coupled to nanoantennas has been studied \cite{3}. The asymmetric coupling, in general, does not result in higher radiation \cite{3}. The break of the symmetry in a resonator e.g. via its asymmetric excitation results in so-called asymmetric modes, e.g. magnetic modes in an initially electric resonator \cite{4}. The radiation of symmetric and asymmetric modes may interfere either constructively or destructively.

\textbf{Page 1:}

All the text of this page is irrelevant and repeats commonplaces of electromagnetism.  The last sentence: "\textit{When the symmetry of the transmission line structure of Fig. 1(b) is broken by opening the wire at its ends [Fig. 1(c)] the translational symmetry of the electric field is broken due to rotation of electric field lines as illustrated in Fig. 1(d) resulting in generation of electromagnetic radiation}" is wrong. In both Figs. 1(b) (diverging two-wire line) and 1 (c) (symmetric dipole) the radiating systems are pretty symmetric. The translational symmetry of the transmission line is irrelevant. If in Fig. 1(b) the divergent line is replaced by the convergent one the wave leakage does not arise. If the dipole arm in Fig. 1(c) is  $\lambda/2$-long the radiation does not arise. The asymmetry of the force lines (in the map of the electric field vector E shown in Fig. 1(d)) is interpreted wrongly. It has nothing to do with the translational symmetry,   it results from the known termination effect \cite{6,7} - radiation of the two-wire line termination interfering with the radiation of the dipole. Even an open-end two-wire line without a dipole radiates \cite{7}.

\textbf{Page 2:}

1. "\textit{In the case of a parallel two wire transmission line, excited by a time varying voltage source… the net magnetic field is  $\mu_{0}(I_{2}-I_{1}$)/2r.}" -- This speculation is irrelevant because is based on formulas of the magneto-statics. For time-varying currents, it does not take into account the phase shift between them, which is relevant for the radiation of a two-wire line.

2. "\textit{The electric field at a distance r outside the upper wire is… which also remains zero}." -- This speculation is fully wrong, because the non-numbered formula, the authors attribute to the electric field of a charged wire, in fact refers to the ideal parallel-plate capacitor. Applied to a wire of finite thickness, it results in the non-physical conclusion that the E field of a thin wire does not depend on the distance and its force lines are parallel. The result violates the energy conservation, and the isotropy of the space. Therefore, the following claims on the role of the asymmetry are wrong.

3. "\textit{In the context of a two wire transmission line, the symmetry of the structure is broken with wires being open ended as shown in Figs. 1(b) and 1(c)… which results in radiation}" -- The same wrong statement on the asymmetry of a radiator resulting in enhanced radiation.

4.  "\textit{…resulting in non-conservation of current and charge in the localized region of the radiating element at that particular instant of time which is the key mechanism of radiation}" -- Again, "\textit{conservation of current}" claimed by the authors as something known is, in fact, mythic.  As to "\textit{conservation of charge in the localized region}", also claimed by the authors as something known, it is meaningless as such. In fact, the law of charge conservation, called continuity equation, relates the variation of the charge in the localized region and the current through the surface of this region. There are no two separate laws of charge conservation and current conservation.

\textbf{Page 3:}

"\textit{…thermal interaction which allows selective transfer of energy from its excitation end to the ground terminal}". -- This is wrong. Thermal interactions (namely a conductive heat exchange, discussed in this paragraph) do not allow any selectivity. On the contrary, the heat exchange works against any selectivity because it increases the entropy in the system (the so-called 2d law of thermodynamics).

\textbf{Page 4:}

1. "\textit{However, some of the fundamental questions related to physics of DRAs remain unclear… they do not have any physical reality [11].} -- All this text except formulas (3,4) is wrong. The concept of polarization currents is taught on the Bachelor level to physicists, radio engineers and opticians ("Electrodynamics of continuous media", "Optics of dielectric media", "Antennas", etc.) In accordance to this concept, the radiation calculates similarly for conductivity currents, currents created by free moving charges and polarization currents. They all emit the dipole radiation described by the same formulas (3,4). If one knows the distribution of the electric field $\mathbf{E}$ in the resonator one knows the distribution of polarization currents whose density $\mathbf{J}$ is proportional to $\mathbf{E}$ with the coefficient $i\omega(\varepsilon-\varepsilon_0)$. Dipole moment of any elementary volume dV of the resonator is equal $\mathbf{J}$dV. Therefore the radiation of any dielectric resonator can be obtained via the volume integration of formulas (3,4) yielded to the surface integration (see e.g. in \cite{2}). As to physical reality of artificial magnetic conductor disclaimed by the authors (and not disclaimed in their Ref. [11] i.e. this reference is also wrong), there is a body of literature on them, reviewed e.g. in \cite{8}.

2. "\textit{The working of DRAs can be understood only in the context of symmetry breaking of dielectric resonators… resulting in an antennalike behavior}" -- Here the wrong statements on the working of DRA and the role of their asymmetry repeat. The effect called "field enhancement under symmetry breaking" and claimed as that accompanying the radiation is, in fact, as irrelevant for the radiation as the asymmetry of the radiator.

\textbf{Page 5:}

"Un\textit{der symmetric connection of the filter… radiation efficiency increased}." -- The authors mix here the asymmetry and the capacitive coupling. Their antenna is in fact the well-known asymmetric dipole, called monopole, one arm of which gives the main contribution into radiation. The radiator of a monopole can be a piece of cable deprived of a metal shield or it can be a metal plate, connected to one wire of the two-wire transmission line conductively and to another wire - inductively or capacitively. Here the metal plate - chassis - is radiating arm, the so-called SAW device operates as a capacitive coupler. In fact, there is no need to use this SAW device at all. The optimal coupler represents a simple metal strip \cite{9}, and there is no reason to use more complicated and expensive devices for the same purpose. The monopole is not better or worse than the symmetric dipole of the same total size, as a radiator. A symmetric dipole can be either strongly radiating or non-radiating depending on the arm length. The non-radiating regime holds if the arms have the length  $\lambda/2$. The monopole can be non-radiating as well - e.g. if its radiating arm has the effective length $\lambda$ \cite{6}. The choice of the monopole instead of a symmetric dipole in the case of a handset antenna is dictated by practical limitations \cite{9}.

\textbf{Page 6:}

1."\textit{The efficiency in all these measurements drops significantly when both the electrodes are excited in a symmetric manner; i.e., input and input ground electrodes are connected to the voltage source}." -- This assertion is trivial and therefore irrelevant. Definitely, if the transmission line is shortcut by a chassis it radiates weakly (only due to its imperfectness). The capacitive connection of the chassis to one feeding wire, whereas the connection to another wire is ohmic, transforms the chassis from a simple shortcut into an arm of the monopole antenna.

2. "\textit{We have shown that… fabrication technology}" -- Here the main wrong claim repeats again and concludes the main paper body.

The authors speculate that some of the fundamental questions related to radiation of DRAs remain unclear and that magnetic currents do not have any physical reality.
On the contrary, Maxwell's equations explicitly demonstrate that a source of the radiation can be either conductivity currents, or currents created by free moving charges, or polarization currents induced in dielectrics. The latter one can be obtained from the known distribution of the electric field $\bf E$ inside the resonator. Based on the field equivalence principle, the radiation of any dielectric resonator can be found either via the volume integration of formulas (3,4) of~\cite{PRL}, or, equivalently, by  surface integration of effective magnetic currents~\cite{2}. Note, that that there is a body of literature on artificial magnetic conductors, see review e.g. in~\cite{8}.

Importantly, no geometrical asymmetry is required to get radiation from dielectric structures. Recently, radiation of optical resonators symmetrically or asymmetrically coupled to nanoantennas has been studied~\cite{3}, including magnetic and electric dipole emission from an spherical silicon nanoparticles~\cite{MagneticLight}. Asymmetric coupling, in general, does not produce stronger radiation; it might lead to the excitation of asymmetric modes, e.g. magnetic modes in an initially electric resonator~\cite{4}.

The claim on the higher efficiency of asymmetric radiators is grounded in~\cite{PRL} not only on the misinterpreted experiment. It is also based on some irrelevant formulas. That for the magnetic field is not applicable to time-varying currents. That for the electric field is not applicable to wires (and, in fact, describes the field in a parallel-plate capacitor). As result, the authors arrive to their misleading claim which, to our opinion, contradicts to both theory and practice and may harm to unexperienced readers.

\bigskip

C. Simovski$^{1,*}$, A. Miroshnichenko$^2$, P. Belov$^3$, and A. Krasnok$^3$

{\small $^1$Aalto University, FI-07615, Finland}

{\small$^2$Australian National University, Canberra, ACT 0200, Australia}

{\small$^3$ITMO University,  St.~Petersburg 197101, Russia}

\medskip
{\small$^{*}$konstantin.simovski@aalto.fi}

\end{document}